\documentclass[aps,prd,reprint,preprintnumbers]{revtex4-2}
\usepackage{graphicx}
\usepackage{amsmath,amssymb,braket,bm}

\begin{document}
\title{Clear correlation between monopoles
and the chiral condensate in SU(3) QCD}
\author{Hiroki~Ohata}
\affiliation{Yukawa Institute for Theoretical Physics, Kyoto University, Kyoto 606-8502, Japan}
\author{Hideo~Suganuma}
\affiliation{Department of Physics, Kyoto University, Kitashirakawaoiwake, Sakyo, Kyoto 606-8502, Japan}
\date{\today}
\begin{abstract}
We study spontaneous chiral-symmetry breaking in SU(3) QCD 
in terms of the dual superconductor picture for quark confinement in the maximally Abelian (MA) gauge, 
using lattice QCD Monte Carlo simulations with 
four different lattices of 
$16^4$, $24^4$, $24^3\times 6$ at $\beta=6.0$ (i.e., the spacing $a \simeq$ 0.1 fm), and
$32^4$ at $\beta=6.2$ (i.e., $a \simeq$ 0.075 fm), 
at the quenched level. 
First, in the confinement phase, we find 
Abelian dominance and monopole dominance in the MA gauge 
for the chiral condensate in the chiral limit,
using the two different methods of 
i) the Banks-Casher relation with the Dirac eigenvalue density   
and ii) finite quark-mass calculations with the quark propagator and its chiral extrapolation. 
In the high-temperature deconfined phase, 
the chiral restoration is observed
also for the Abelian and the monopole sectors. 
Second, we investigate local correlation between 
the chiral condensate and monopoles, which 
topologically appear in the MA gauge. 
We find that the chiral condensate locally takes a quite large value near monopoles. 
As an interesting possibility, 
the strong magnetic field around monopoles 
is responsible to chiral symmetry breaking in QCD, 
similarly to the magnetic catalysis.
\end{abstract}

\maketitle

\section{Introduction}
Since quantum chromodynamics (QCD) was established 
as the fundamental theory of strong interaction in 1970s, 
to understand its nonperturbative properties has been one of the 
most difficult central problems in theoretical physics for about a half century.
In particular, QCD exhibits two outstanding nonperturbative phenomena 
of quark confinement and spontaneous chiral-symmetry breaking 
in its low-energy region, many physicists have tried 
to clarify these phenomena and their relation directly from QCD, 
but this is still an unsolved important issue in the particle physics. 

Chiral symmetry breaking in QCD is categorized 
as well-known spontaneous symmetry breaking, 
which widely appears in various fields in physics, 
and is an important phenomenon 
relating to dynamical quark-mass generation ~\cite{Nambu:1961tp,Higashijima:1983gx}.
Apart from the dark matter, about 99\% of the matter mass of our Universe 
originates from chiral symmetry breaking, 
because the Higgs-origin mass is just a small mass of u, d current quarks, electrons, and neutrinos~\cite{Zyla:2020zbs}. 
The order parameter of chiral symmetry breaking 
is the chiral condensate $\langle \bar qq \rangle$, 
and it is directly related to low-lying Dirac modes, 
via the Banks-Casher relation~\cite{Banks:1979yr}.

In contrast, color confinement is a fairly unique phenomenon 
peculiar in QCD, 
and quark confinement is characterized by 
the linear inter-quark potential.
As for the confinement mechanism,  
the dual superconductor picture based on color-magnetic monopole condensation 
was proposed by Nambu, 't~Hooft, and Mandelstam 
as a typical plausible physical scenario~\cite{Nambu:1974zg,tHooft:1975krp,Mandelstam:1974pi}.
In lattice QCD, by taking the maximally Abelian (MA) gauge~\cite{Kronfeld:1987ri}, 
this dual superconductor scenario has been investigated
in terms of Abelian dominance, i.e., dominant role of the Abelian sector
~\cite{Suzuki:1989gp,Amemiya:1998jz,Sakumichi:2014xpa,Ohata:2020dgn}.
and the relevant role of monopoles~ \cite{Stack:1994wm,Bakker:1997rk,Ichie:1998ns}.
 
The relation between confinement 
and chiral symmetry breaking 
is not yet clarified directly from QCD.
While a strong correlation between confinement and chiral symmetry breaking 
has been suggested by 
almost coincidence between deconfinement and chiral-restoration temperatures~\cite{Rothe:2012nt}, 
an lattice QCD analysis based on the Dirac-mode expansion 
indicates some independence of these phenomena~\cite{Doi:2014zea}.

Their correlation has been also suggested in terms of color-magnetic monopoles, 
which topologically appear in QCD in the Abelian gauge~\cite{tHooft:1981bkw}. 
In the dual Ginzburg-Landau theory, 
the monopole condensate is responsible to chiral symmetry breaking
as well as quark confinement~\cite{Suganuma:1993ps}.
Also in SU(2) lattice QCD, Miyamura and Woloshyn showed 
Abelian dominance~\cite{Miyamura:1995xn,Woloshyn:1994rv} and 
monopole dominance~\cite{Miyamura:1995xn,Lee:1995ac} 
for chiral symmetry breaking. 
In fact, by removing the monopoles from the QCD vacuum, 
confinement and chiral symmetry breaking are simultaneously lost.
In SU(3) lattice QCD with a $8^3 \times 4$ lattice,
Thurner et al. showed a local correlation among monopoles,
instantons, and the chiral condensate~\cite{Thurner:1997qx}. 
These studies indicate an important role of the monopoles to both phenomena, 
and thus these two phenomena might be related via the monopole.
However, most of the pioneering lattice works were done in SU(2) lattice QCD
or done on a small lattice~\cite{Miyamura:1995xn,Woloshyn:1994rv,Lee:1995ac,Thurner:1997qx}.

In this paper, we investigate correlation between chiral symmetry breaking 
and color-magnetic monopoles appearing in the MA gauge 
in SU(3) lattice QCD with large-volume fine lattices at the quenched level. 
Using two different methods, 
we evaluate the chiral condensate in Abelianized QCD and the monopole system, 
extracted from lattice QCD.
We also investigate correlation between the local chiral-condensate value and the monopole location.

\section{Lattice setup and Abelian projection}
We perform SU(3) lattice QCD simulations 
at the quenched level with the standard plaquette action~\cite{Rothe:2012nt}.
On four-dimensional Euclidean lattices, 
the gauge variable is described as the SU(3) link variable 
$U_{\mu}(s) \equiv e^{iagA_{\mu}(s)} \in {\rm SU(3)}$, 
with the gluon field $A_{\mu}(s) \in {\rm su(3)}$, 
the QCD gauge coupling $g$, and the lattice spacing $a$.
The lattice spacing $a$ is determined 
so as to reproduce the string tension 
$\sigma$=0.89 GeV/fm~\cite{Sakumichi:2014xpa}.

In this work, we use four different lattices with 
the size and the lattice parameter $\beta \equiv 6/g^2$:
\begin{itemize}
\item[(a)] 
$16^3 \times 16$ 
and $\beta = 6.0$ 
(i.e., $a \simeq$ 0.1 fm),
\item[(b)]
$24^3 \times 24$
and $\beta = 6.0$, 
\item[(c)]
$32^3 \times 32$ 
 and $\beta = 6.2$
 (i.e., $a \simeq$ 0.075 fm), 
\item[(d)] 
$24^3 \times 6$ and $\beta = 6.0$.
\end{itemize}
From the first and the second lattices, the finite volume effect can be checked.
Since the second and the third lattices have almost the same physical volume, 
the finite lattice-spacing effect can be also checked. 
The last one exhibits the high-temperature deconfined phase 
at $T\simeq$ 330 MeV above the critical temperature. In each direction, the periodic boundary condition is imposed for link variables, and the anti-periodic boundary condition for quarks, 
which realizes the finite temperature system for the last lattice.

Hereafter, we take the lattice unit $a=1$.
Using the pseudo-heat-bath algorithm, 
we generate 300, 100, 100, and 300 gauge configurations for the lattices of (a), (b), (c), and (d), respectively.  
All of gauge configurations are taken every 500 sweeps after a thermalization of 5,000 sweeps.
We use the jackknife method for the statistical error estimate.

Using the Cartan subalgebra $\vec{H} \equiv (T_3,T_8)$ of SU(3), 
the MA gauge fixing is defined so as to maximize 
\begin{align}
  R_{\rm MA}[U_{\mu}(s)] \equiv& \sum_s \sum_{\mu=1}^4 {\rm tr} 
  \left( U_{\mu}^\dag(s) \vec{H} U_{\mu}(s) \vec{H} \right) 
  \notag \\
  =& \sum_{s} \sum_{\mu=1}^4 
  \left( 1 - \frac{1}{2}\sum_{i \neq j} 
\left| U_{\mu}(s)_{ij} \right|^2 \right) \label{eq:Abelian}
\end{align}
under the SU(3) gauge transformation, and thus 
this gauge fixing suppresses all the off-diagonal fluctuation 
of the SU(3) field $U_\mu(s)$. 
In the MA gauge, the SU(3) gauge group is partially fixed remaining 
its maximal torus subgroup ${\rm U(1)}_3 \times {\rm U(1)}_8$,
and QCD is reduced into an Abelian gauge theory like the non-Abelian Higgs theory.

In this work, the MA gauge fixing is performed with the stopping criterion 
that the deviation $\Delta R_{\mathrm{MA}} / \left( 4V^4 \right)$ 
becomes smaller than $10^{-5}$ in 100 iterations.

From the SU(3) field $U_{\mu}^{\rm MA}(s) \in {\rm SU(3)}$ in the MA gauge,  
the Abelian field is defined as 
\begin{equation}
  u_{\mu}(s) = e^{i\vec \theta \cdot \vec H}
  =\mathrm{diag}\left( e^{i\theta_{\mu}^{1}(s)}, e^{i\theta_{\mu}^{2}(s)},
  e^{i\theta_{\mu}^{3}(s)} \right) 
  \in \mathrm{U}(1)^2 \label{eq:Abelian_field} 
\end{equation}
with the constraint $\sum_{i=1}^{3} \theta_{\mu}^{i}(s) = 0 \left( \mathrm{mod}~2\pi \right)$,
by maximizing the overlap
\begin{equation}
  R_{\rm Abel} \equiv 
  \frac{1}{3} {\rm Re \, tr} \left\{ U_{\mu}^{\rm MA}(s) u_{\mu}^\dag(s) \right\}
  \in \left[ -\frac{1}{2}, 1 \right],
\end{equation}
so that the distance between $u_{\mu}(s)$ and $U_{\mu}^{\rm MA}(s)$
becomes the smallest in the SU(3) manifold. 

The Abelian projection is defined 
by the replacement of SU(3) fields $U_\mu(s)$ 
by Abelian fields $u_\mu(s)$ for each gauge configuration, i.e., 
$O[U_\mu(s)] \rightarrow O[u_\mu(s)]$ for QCD operators. 
In this way, Abelian-projected QCD is extracted from SU(3) QCD.
The case of $\langle O[U_\mu(s)] \rangle \simeq \langle O[u_\mu(s)] \rangle$ 
is called ``Abelian dominance'' for the operator $O$. 

\section{Monopoles in QCD}
Now, let us consider the Abelian plaquette variable, 
\begin{eqnarray}
u_{\mu\nu}(s) 
&\equiv&
u_{\mu}(s) u_{\nu}(s+\hat \mu) u_{\mu}^\dagger(s+\hat \nu)  u_{\nu}^\dagger(s) 
=e^{i\theta_{\mu\nu}(s)} \cr
&=&{\rm diag}(e^{i\theta_{\mu\nu}^{1}(s)},e^{i\theta_{\mu\nu}^{2}(s)}, e^{i\theta_{\mu\nu}^{3}(s)}) \in {\rm U(1)}^2,
\end{eqnarray}
where $\hat \mu$ is the $\mu$-directed unit vector in the lattice unit.
The Abelian field strength $\theta_{\mu \nu}^{i}(s)$ ($i=1,2,3$) is 
the principal value of the exponent in $u_{\mu\nu}(s)$, and is defined as 
\begin{align}
  \partial_{\mu} \theta_{\nu}^{i}(s) - \partial_{\nu}\theta_{\mu}^{i}(s) 
  &= \theta_{\mu\nu}^{i}(s) - 2\pi n_{\mu\nu}^{i}(s), \notag \\
  -\pi \le \theta_{\mu\nu}^{i}(s) &< \pi, \quad n_{\mu\nu}^{i}(s) \in \mathbb{Z},
\end{align}
with the forward derivative $\partial_{\mu}$. 
Here, $\theta_{\mu\nu}^{i}(s)$ is ${\rm U(1)}^2$ gauge invariant 
and corresponds to the regular continuum Abelian field strength 
as $a \rightarrow 0$, while 
$n_{\mu\nu}^{i}(s)$ corresponds to 
the singular gauge-variant Dirac string~\cite{DeGrand:1980eq}.

The electric current $j_\mu^i$ and the monopole current $k_\mu^i$ are 
defined from the Abelian field strength $\theta_{\mu\nu}^i$, 
\begin{eqnarray}
j_\nu^i(s) &\equiv& \partial_\mu^{\prime} \theta_{\mu\nu}^i(s), 
\\
k_\nu^i(s) &\equiv& \partial_\mu \tilde{\theta}_{\mu\nu}^i(s)/2\pi
=\partial_\mu \tilde{n}_{\mu\nu}^i \in \mathbb{Z},
\end{eqnarray}
where $\partial_{\mu}^{\prime}$ is the backward derivative,
and $\tilde{\theta}_{\mu\nu}$ denotes the dual tensor of 
$\tilde{\theta}_{\mu\nu} \equiv \frac12 \epsilon_{\mu\nu\alpha\beta}{\theta}_{\alpha\beta}$.
Both electric and monopole currents are $\mathrm{U}(1)^2$ gauge invariant, 
according to $\mathrm{U}(1)^2$ gauge invariance of ${\theta}_{\mu\nu}^i(s)$.
In the lattice formalism, 
$k_{\mu}^{i}(s)$ is located at the dual lattice 
$L_{\rm dual}^4$
of $s^\alpha+1/2$, flowing in $\mu$ direction~\cite{Ichie:1998ns}.
Hereafter, we will omit the color index $i$ as appropriate. 

Abelian-projected QCD thus includes both 
electric current $j_\mu$ and monopole current $k_\mu$, 
and can be decomposed into 
the ``photon part'' which only includes $j_\mu$ 
and the ``monopole part'' which only includes $k_\mu$ approximately, as follows.

First, we consider the photon part satisfying
\begin{align}
&\theta_{\mu \nu}^{\rm Ph}
\equiv {\rm mod}_{2\pi}(\partial \wedge \theta^{\rm Ph})_{\mu\nu}, \label{eq:Ph} \\
&\partial_\mu^{\prime} \theta_{\mu \nu}^{\rm Ph}=j_\nu, \quad 
\partial_\mu \tilde{\theta}_{\mu \nu}^{\rm Ph}=0.
\end{align}
Here, we denote by ${\rm mod}_{2\pi}$ the principal value in  $[-\pi,\pi)$.
From $\partial_\mu \tilde{\theta}_{\mu \nu}^{\rm Ph}=0$, one can set  
$\theta_{\mu \nu}^{\rm Ph}
=(\partial \wedge \theta^{\rm Ph})_{\mu\nu}$
and then 
$\partial_\mu^{\prime} (\partial \wedge \theta^{\rm Ph})_{\mu\nu}
=\partial^2 \theta_\nu^{\rm Ph} - 
\partial_\mu^{\prime} \partial_\nu \theta_\mu^{\rm Ph}
=j_\nu$.
In the Landau gauge $\partial_\mu^{\prime} \theta_\mu^{\rm Ph}=0$,
the photon part $\theta_\nu^{\rm Ph}$ can be derived from the electric current $j_\nu$, 
\begin{eqnarray}
\partial^2 \theta_\nu^{\rm Ph}=j_\nu, \quad
\theta_\nu^{\rm Ph}=\frac{1}{\partial^2} j_\nu. 
\end{eqnarray}
Therefore, we here define the photon part $\theta_\nu^{\rm Ph}$ by 
\begin{eqnarray}
\quad \theta_\nu^{\rm Ph}(s)
\equiv \sum_{s'} \langle s|\frac{1}{\partial^2}|s'\rangle j_\nu(s'), 
\end{eqnarray}
using the inverse d'Alembertian 
on the lattice~\cite{Ichie:1998ns},
\begin{eqnarray}
\langle s|\frac{1}{\partial^2}|s'\rangle =f((s-s')a)
\end{eqnarray}
with
\begin{eqnarray}
f(s a) \equiv
-\frac{a^2}{4}\int_{-\frac{\pi}{a}}^{\frac{\pi}{a}}
\frac{d^4p}{(2\pi)^4}\frac{e^{-ip_\alpha s_\alpha a}}
{\sum_{\mu=1}^4\sin^2(p_\mu a/2)}.
\label{eq:inverse}
\end{eqnarray}
Here, we have explicitly written the lattice spacing $a$.
Note that this function satisfies
\begin{eqnarray}
\partial^2 f(s a) 
&\equiv&\partial_\mu \partial'_\mu f(s a) 
\cr
&=&\frac{1}{a^2}\sum_{\mu=1}^4 
[f(sa+\hat \mu)+f(sa-\hat \mu)-2f(sa)]
\cr
&=&\int_{-\frac{\pi}{a}}^{\frac{\pi}{a}}
\frac{d^4p}{(2\pi)^4}e^{-ip_\alpha s_\alpha a},
\end{eqnarray}
which goes to the four-dimensional delta function 
in the continuum limit $a \rightarrow 0$.
On finite-size lattices, the momentum integral in 
Eq.(\ref{eq:inverse}) becomes a discretized sum 
over the momentum-space lattice.

The monopole part $\theta_\mu^{\rm Mo}(s)$ is defined as
\begin{eqnarray}
\theta_\mu^{\rm Mo}(s) \equiv \theta_\mu(s)-\theta_\mu^{\rm Ph}(s),
\end{eqnarray}
and approximately satisfies
\begin{align}
&\theta_{\mu \nu}^{\rm Mo}
\equiv {\rm mod}_{2\pi}(\partial \wedge \theta^{\rm Mo})_{\mu\nu}, \label{eq:Mo} \\
&\partial_\mu^{\prime} \theta_{\mu \nu}^{\rm Mo} \simeq 0, \quad
\partial_\mu \tilde{\theta}_{\mu \nu}^{\rm Mo} \simeq k_\nu.
\end{align}

In this way, in Abelian-projected QCD, 
the contributions from the electric current $j_\mu$ and 
the magnetic current $k_\mu$ can be well separated into 
the photon part $\theta_\mu^{\rm Ph}$ 
and the monopole part $\theta_\mu^{\rm Mo}$, respectively.

In Table~\ref{table:density}, we show 
the monopole density $\rho_{\mathrm{M}}$ and the electric-current density $\rho_{\mathrm{E}}$ defined as
\begin{eqnarray}
\rho_{\mathrm{M}} &\equiv \frac{1}{3V} \sum_{i=1}^{3}\sum_{s,\mu}\left| k_{\mu}^{i}(s) \right|, \\
\rho_{\mathrm{E}} &\equiv \frac{1}{3V} \sum_{i=1}^{3}\sum_{s,\mu} \left| j_{\mu}^{i}(s) \right|
\end{eqnarray}
for Abelian-projected QCD, monopole and photon parts, respectively.

\begin{table}[htb]
  \centering
  \begin{tabular}{cccc}
    \hline \hline
    Lattice & Field sector & Monopole density & Electric density \\ 
    \hline
     $V = 16^3 \times 16$ & Abel & $2.95(2) \times 10^{-2}$ & 6.932(1) \\
     $\beta = 6.0$ & Monopole & $2.37(2) \times 10^{-2}$ & $0.0967(5)$ \\
     & Photon  & $1.39(3) \times 10^{-4}$ & 6.906(1) \\
     \hline 
     $V = 24^3 \times 24$ & Abel & $2.94(1) \times 10^{-2}$ & 6.9307(7) \\  
     $\beta = 6.0$ & Monopole & $2.35(1) \times 10^{-2}$ & $0.0964(4)$ \\ 
     & Photon   & $1.39(2) \times 10^{-4}$ & 6.9048(7) \\ 
    \hline
     $V = 32^3 \times 32$ & Abel & $1.065(5) \times 10^{-2}$ & 6.5190(5) \\  
     $\beta = 6.2$ & Monopole & $0.842(5) \times 10^{-2}$ & $0.0338(1)$ \\ 
     & Photon   & $4.01(9) \times 10^{-5}$ & 6.5100(4) \\ 
    \hline
     $V = 24^3 \times 6$  & Abel & $1.720(9) \times 10^{-2}$ & 6.8760(8) \\  
     $\beta = 6.0$ & Monopole & $1.229(8) \times 10^{-2}$ & $0.0654(3)$ \\ 
     & Photon   & $9.2(2) \times 10^{-5}$ & 6.8598(7) \\ 
    \hline \hline
  \end{tabular}
  \caption{The monopole density $\rho_{\mathrm{M}}$ and
  the electric-current density $\rho_{\mathrm{E}}$ 
  for Abelian-projected QCD, monopole and photon parts.}
  \label{table:density}
\end{table}

Using the monopole and the photon link-variables, 
\begin{eqnarray}
u_\mu^{\rm Mo}(s) \equiv e^{i \theta_\mu^{\rm Mo}(s)}
\in {\rm U(1)}^2, \\
u_\mu^{\rm Ph}(s) \equiv e^{i \theta_\mu^{\rm Ph}(s)} 
\in {\rm U(1)}^2, 
\end{eqnarray}
monopole and photon projection are defined by
the replacement of $\{u_\mu(s)\} \rightarrow \{u_\mu^{\rm Mo}(s)\}, 
\{u_\mu^{\rm Ph}(s)\}$.
%
The dominant role of the monopole part is called 
``monopole dominance,'' 
and monopole dominance has been observed for quark confinement 
in lattice QCD~\cite{Stack:1994wm}.

\section{Chiral condensate}
First, we study Abelian dominance and
monopole dominance for the chiral condensate in the chiral limit, 
using the Kogut-Susskind (KS) fermion~\cite{Rothe:2012nt} 
for quarks in SU(3) lattice QCD. 

Mathematically, the chiral condensate $\langle \bar qq \rangle$ in the chiral limit 
is directly related to the low-lying Dirac eigenvalue density $\rho(0)$ 
through the Banks-Casher relation~\cite{Banks:1979yr},  
\begin{eqnarray}
\langle\bar{q}q\rangle=-\lim_{m\to0}\lim_{V\to\infty}\pi \rho(0).
\end{eqnarray}
The Dirac eigenvalue density $\rho(\lambda)$ is defined as
\begin{eqnarray}
\rho(\lambda)\equiv \frac{1}{V}\sum_n\langle\delta(\lambda-\lambda_n)\rangle, 
\quad
\gamma_\mu D_\mu |n \rangle =i \lambda_n |n \rangle
\end{eqnarray}
with the space-time volume $V$. 

\begin{figure*}[htb]
  \centering
  \includegraphics[width=\textwidth]{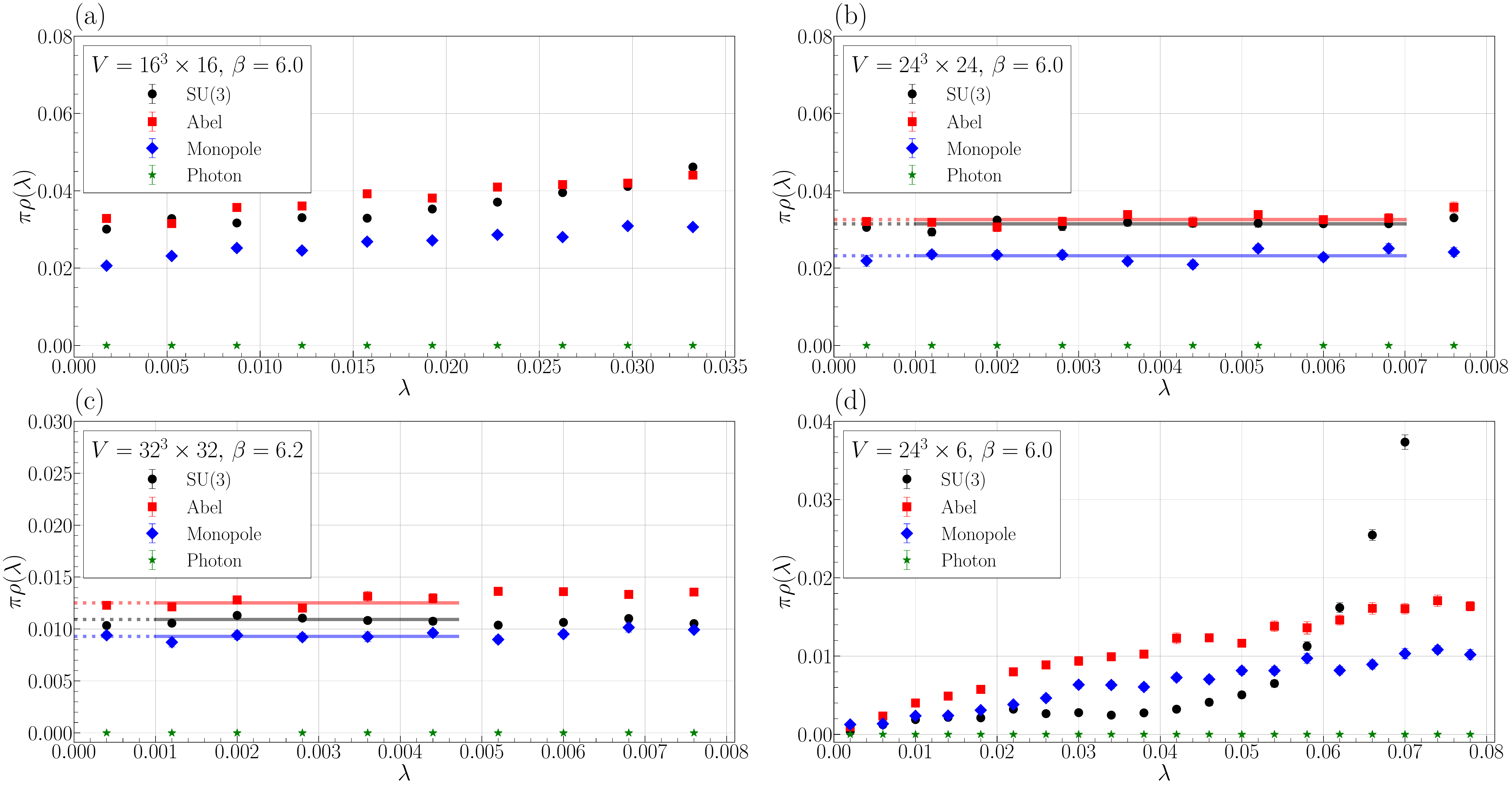}
  \caption{The Dirac eigenvalue densities $\rho(\lambda)$ 
  for SU(3) QCD, Abelian-projected QCD, monopole and photon sectors,  
  as functions of the Dirac eigenvalue $\lambda$, for the four types of different lattices: 
  (a) $16^3 \times 16$ at $\beta$=6.0, 
  (b) $24^3 \times 24$ at $\beta$=6.0, 
  (c) $32^3 \times 32$ at $\beta$=6.2, 
  and (d) $24^3\times 6$ at $\beta$=6.0.
  The line is the best fit with a constant for the low-lying Dirac eigenvalue density.}
  \label{fig:diracmode}
\end{figure*}

For the KS fermion, 
the Dirac operator $\gamma_\mu D_\mu$ becomes
$\eta_\mu D_\mu$ with the staggered phase 
$\eta_\mu(s)\equiv (-1)^{s_1+\cdots+s_{\mu-1}}$ 
$(\mu \geq 2)$ 
with $\eta_1(s)\equiv 1$.
The KS Dirac operator takes the explicit form of 
\begin{eqnarray}
(\eta_\mu D_\mu)_{ss'}&=&\frac{1}{2}\sum_{\mu=1}^{4}\eta_\mu(s)
[U_\mu(s)\delta_{s+\hat{\mu},s'}
-U_{-\mu}(s) \delta_{s-\hat{\mu},s'}]\cr
&=&\frac{1}{2}\sum_{\mu=1}^{4}\sum_{\pm}
\pm \eta_\mu(s)
U_{\pm \mu}(s)\delta_{s \pm \hat{\mu},s'},
\end{eqnarray}
and hence the Dirac eigenvalue $\lambda_n$ 
is obtained from 
\begin{eqnarray}
\frac{1}{2}\sum_{\mu=1}^4 \sum_{\pm} 
\pm \eta_\mu(s) U_{\pm\mu}(s) \chi_n(s\pm\hat \mu)
=i\lambda_n\chi_n(s). 
\end{eqnarray}
Here, the quark field $q^\alpha(x)$ is described by 
a spinless Grassmann variable $\chi(x)$, 
and the chiral condensate per flavor is given as  
$\langle \bar qq\rangle=\langle \bar \chi \chi\rangle$/4 in the continuum limit.

For the four types of different lattices, we show in Fig.~\ref{fig:diracmode} 
the Dirac eigenvalue densities $\rho(\lambda)$ 
for SU(3) QCD, Abelian-projected QCD, monopole and photon sectors, extracted from lattice QCD in the MA gauge 
as functions of the Dirac eigenvalue $\lambda$.
Figure~\ref{fig:diracmode} (a), (b), and (c) are almost zero-temperature results 
in the confined phase, and Fig.~\ref{fig:diracmode} (d) exhibits the high-temperature deconfined phase above the critical temperature.

For all the four lattices, 
we find that the low-lying Dirac eigenvalue density $\rho(0)$ in Abelian-projected QCD takes approximately the same value in SU(3) QCD, 
which means Abelian dominance for the chiral condensate in the chiral limit.
For the photon sector, we find no eigenvalues below 0.20, 0.13, 0.098, and 0.27
in 10 configurations for the four different lattices of (a), (b), (c), and (d), respectively, 
and conclude that $\rho(0)$ in the photon sector is exactly zero.
On the other hand, $\rho(0)$ in the monopole part is close to that in SU(3) QCD, which means monopole dominance for the chiral condensate in the chiral limit.
Also, in the high-temperature deconfined phase of Fig.~\ref{fig:diracmode}(d), 
one finds $\rho(0) \simeq 0$ for all the sectors, 
which physically means chiral restoration. 

Next, we calculate the chiral condensate in a different way using the quark propagator. 
Here, we adopt the KS fermion with the bare quark mass $m$, 
and consider the chiral extrapolation of $m \rightarrow 0$. 

\begin{figure*}[htb]
  \centering
  \includegraphics[width=\textwidth]{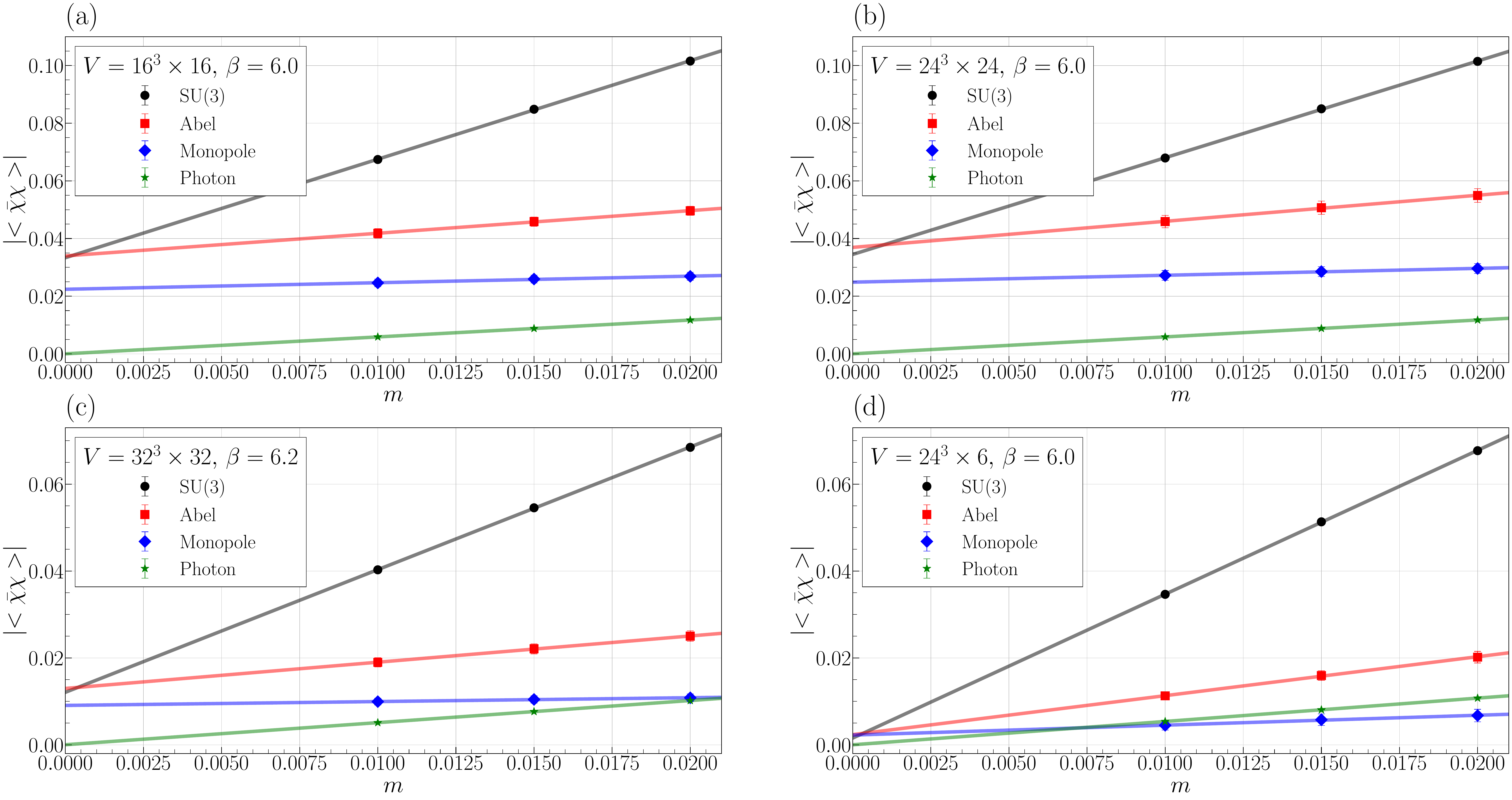}
  \caption{The chiral condensates for SU(3) QCD, Abelian-projected QCD, monopole and photon sectors,
as functions of the bare quark mass $m$ in the lattice unit, for the four types of different lattices: 
(a) $16^3 \times 16$ at $\beta$=6.0, 
(b) $24^3 \times 24$ at $\beta$=6.0, 
(c) $32^3 \times 32$ at $\beta$=6.2, 
and (d) $24^3\times 6$ at $\beta$=6.0.
The solid line is the best fit with a linear function.
}
  \label{fig:chiral_expo}
\end{figure*}

For the gauge-field ensemble of $U=\{U_\mu(s)\}$, 
the Euclidean KS fermion propagator is given 
by the inverse matrix,
\begin{eqnarray}
G_U^{ij}(x,y) &\equiv& \langle {\chi}^i(x) \bar \chi^j(y) \rangle_U \cr
&=&\langle x,i|\left(
\frac{1}{\eta_\mu D_\mu[U]+m}\right)|y,j\rangle, 
\end{eqnarray}
with the color index $i$ and $j$.
The propagator is calculated 
by solving the large-scale linear equation 
with a point source. 
Using the propagator for  
the gauge-field ensemble $\{U_\mu(s)\}$, $\{u_\mu(s)\}$, $\{u_\mu^{\rm Mo}(s)\}$, and $\{u_\mu^{\rm Ph}(s)\}$, 
we calculate the local chiral condensate 
\begin{eqnarray}
\langle \bar \chi(x)\chi(x) \rangle_U=-{\rm Tr}\, G_U(x,x)
\end{eqnarray}
for SU(3) QCD, Abelian-projected QCD, 
monopole and photon sectors, respectively. 
Here, we use 100 gauge configurations, and 
calculate the local chiral condensate 
at $2^4$ distant space-time points $x$ 
for each gauge configuration.
In fact, we perform 1,600 times calculations of 
$\langle \bar \chi(x) \chi(x)\rangle_U$ 
for each sector, quark mass $m$, and type of lattice. 
Here, we consider the net chiral condensate by subtracting  
the contribution from the trivial vacuum $U = 1$ as
\begin{eqnarray}
\langle \bar \chi\chi(x) \rangle_U
\equiv 
\langle \bar \chi(x)\chi (x)\rangle_U
-\langle \bar \chi\chi \rangle_{U=1},
\end{eqnarray}
where the subtraction term is exactly zero 
at the chiral limit $m=0$. 
We eventually take its average
over the space-time $x$
and the gauge ensembles $U_1, U_2, ..., U_N$,
\begin{eqnarray}
\langle \bar \chi\chi\rangle
\equiv \sum_{x,i}
\langle \bar \chi\chi (x)\rangle_{U_i}/\sum_{x,i} 1.
\end{eqnarray}

For the four types of different lattices, we show in Fig.~\ref{fig:chiral_expo} 
the chiral condensates 
plotted against the bare quark mass $m$ in the lattice unit,  
for SU(3), Abelian, monopole and photon sectors, 
extracted from lattice QCD in the MA gauge.

For each sector, $m$-dependence of the chiral condensate seems to be linear in this region, and therefore  
we evaluate the chiral condensate in the chiral limit,  
using the linear chiral extrapolation.
Provided that the linear chiral extrapolation is valid, 
Abelian dominance and monopole dominance for the chiral condensate are realized in the chiral limit, 
whereas the photon part has almost no chiral condensate 
in the chiral limit.

These results are consistent with 
the above-mentioned conclusions using 
the Dirac eigenvalue density $\rho(\lambda)$ and the Banks-Casher relation.
In Table~\ref{table:result}, we summarize the chiral condensate values in the chiral limit evaluated from the two different methods for SU(3), Abelian, monopole and photon sectors.

\begin{table}[htb]
  \begin{tabular}{cccc}
    \hline \hline
    Lattice & Field sector & Banks-Casher & Propagator \\
    \hline
    $V = 24^3 \times 24$ 
    & SU(3)    & $3.14(3) \times 10^{-2}$ & $3.45(5) \times 10^{-2}$ \\
    $\beta = 6.0$ 
    & Abel     & $3.26(4) \times 10^{-2}$ & $3.69(5) \times 10^{-2}$ \\
    & Monopole & $2.32(5) \times 10^{-2}$ & $2.49(2) \times 10^{-2}$ \\
    & Photon   & 0.00(0) & $2.5(7) \times 10^{-5}$ \\
    \hline
    $V = 32^3 \times 32$ 
    & SU(3)    & $1.09(1) \times 10^{-2}$ & $1.21(3) \times 10^{-2}$ \\
    $\beta = 6.2$                             
    & Abel     & $1.25(2) \times 10^{-2}$ & $1.30(2) \times 10^{-2}$ \\
    & Monopole & $0.93(1) \times 10^{-2}$ & $0.91(1) \times 10^{-2}$ \\
    & Photon   & 0.00(0) & $2.4(7) \times 10^{-5}$ \\
     \hline \hline
  \end{tabular}
  \caption{The chiral condensate values in the chiral limit evaluated from the two different methods for SU(3) QCD, Abelian-projected QCD, monopole and photon sectors.}
  \label{table:result}
\end{table}

In the presence of bare quark masses of $m = 0.01 - 0.02$, however, 
there appears a significant deviation of the chiral condensates 
between SU(3) and Abelian sectors, 
which quantitatively differs from SU(2) QCD, 
where Abelian dominance is observed at $m = 0.05 - 0.3$~\cite{Woloshyn:1994rv}.
In particular, compared with SU(3) QCD, 
bare-quark mass $m$ dependence of the chiral condensate 
is fairly reduced in Abelian-projected QCD, 
and also in the monopole part. 

As an interesting possibility,
the net chiral condensate 
in the Abelian/monopole sector 
is controlled by quark-mass independent object. 
This might be understood if monopoles are directly 
responsible for chiral symmetry breaking, 
because monopoles have no bare quark mass dependence 
in the quenched approximation.
Then, 
we next examine the correlation  
between the chiral condensate and monopoles 
in more direct manner.

\section{Local correlation}
Second, we study the local correlation between chiral condensate and monopoles, 
by investigating the local chiral condensate around monopoles 
in Abelian-projected QCD 
at each gauge configuration. 
Note that, at each lattice configuration,
the monopoles topologically appear as local objects, 
so that they might locally influence the chiral condensate around them, 
although the translational invariance is recovered by the gauge ensemble average.

\begin{figure}[htb]
  \centering
  \includegraphics[width=0.5\textwidth]{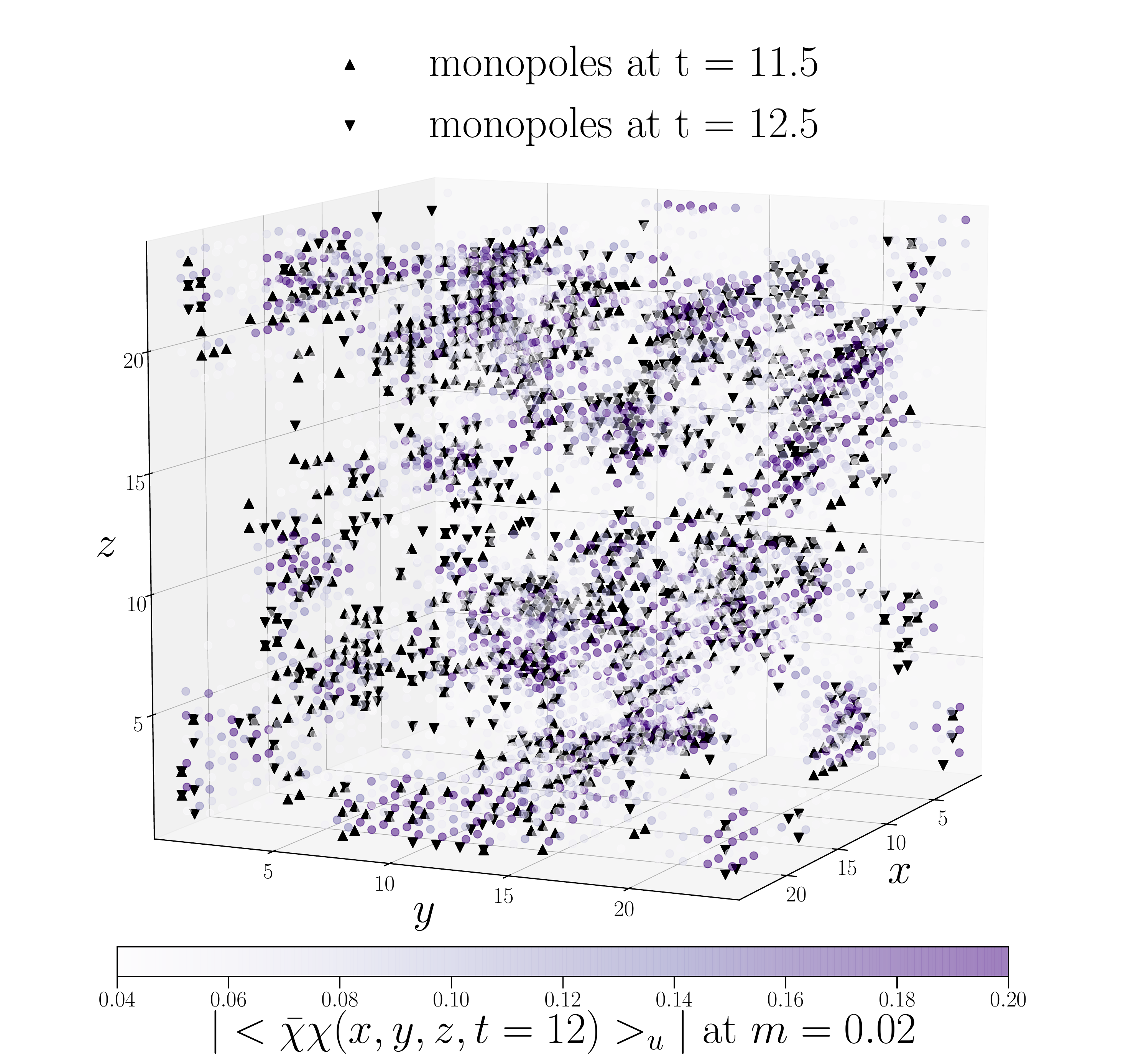}
  \caption{The local chiral condensate at $t = 12$ and monopoles at $t = 11.5, 12.5$ 
  for a typical Abelian configuration of 
  the $24^3 \times 24$ lattice at $\beta$=6.0.
  The bare quark mass is taken as $m = 0.02$.
  The value of the local chiral condensate 
  $|\langle \bar \chi \chi(x)\rangle_u|$ 
  is visualized with the color graduation. 
  Monopoles at $t=11.5$ and $12.5$ are plotted with upper and lower triangles, respectively.}
  \label{fig:3dim_chiral}
\end{figure}

For the visual demonstration, 
we show in Fig.~\ref{fig:3dim_chiral} 
the local chiral condensate 
$\langle \bar \chi \chi(x)\rangle_u$  
and the monopole location 
at all three-dimensional space points at a time slice of $t=12$ 
in a typical Abelian configuration of 
the $24^3 \times 24$ lattice at $\beta$=6.0. 
The bare quark mass is taken as $m = 0.02$.
Here, we show all the monopoles located at $t=11.5, 12.5$ on the dual lattice
$L_{\rm dual}^4$ of $s^\alpha+1/2$. 
The value of the local chiral condensate $|\langle \bar \chi \chi(x)\rangle_u|$  
is visualized with the color graduation. 
(The same dark color is used for $|\langle \bar \chi \chi(x)\rangle_u| \ge 0.20$, and 
no color is used for small $|\langle \bar \chi \chi(x)\rangle_u|<0.04$.)

It is clearly observed that the local chiral condensate in a configuration has  
a large fluctuation and takes quite large values 
in the vicinity of the monopoles. 

Finally, we calculate the 
correlation function between the local chiral condensate 
$\langle \bar \chi \chi(x)\rangle_u$ 
and the local monopole density 
\begin{eqnarray}
\rho_{\mathrm{L}}(s) \equiv \frac{1}{3 \cdot 2^4} 
\sum_{i=1}^{3}
\sum_{s'\in P(s)}
\sum_{\mu=1}^4
\left| k_{\mu}^{i}(s')
 \right|,
\end{eqnarray}
where $P(s)$ denotes the dual lattices in the vicinity of $s$, i.e.,
$P(s)=\left\{s'\in L_{\rm dual}^4 \middle| 
|s'-s|=1 \right\}$ 
with the dual lattice $L_{\rm dual}^4$ of $s^\alpha+1/2$.
For this calculation, we use 
the lattice data of the local chiral condensate and the monopole current 
for 100 gauge configurations, which were used to obtain Fig.~\ref{fig:chiral_expo}. 

\begin{figure*}[htb]
  \centering
  \includegraphics[width=\textwidth]{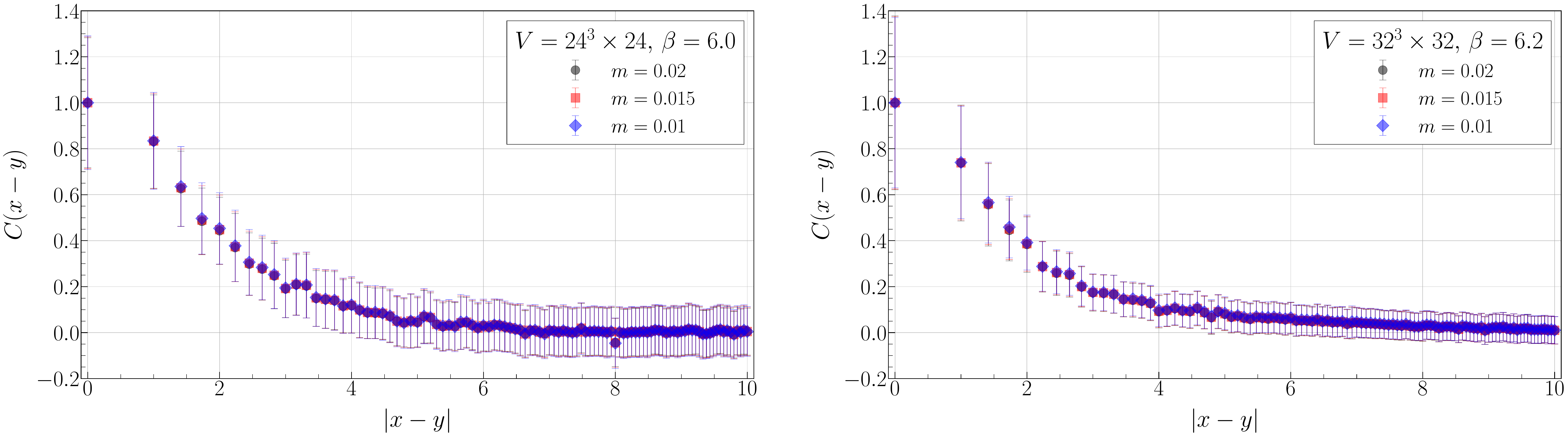}
  \caption{The correlation function $C(x-y)$ between the local chiral condensate 
  $\langle \bar \chi \chi(x)\rangle_u$ 
  and the local monopole density 
  $\rho_{\mathrm{L}}(y)$
  plotted against $|x-y|$ for
(left) $24^3 \times 24$ at $\beta$=6.0 and (right) $32^3 \times 32$ at $\beta$=6.2.
The bare quark masses 
of $m$ = 0.02, 0.015, and 0.01 are used in the lattice unit.
}
  \label{fig:corrfunc}
\end{figure*}

Figure~\ref{fig:corrfunc} shows 
the correlation function $C(x-y)$ between 
the local chiral condensate 
$\langle \bar \chi \chi(x)\rangle_u$ 
and the local monopole density $\rho_{\mathrm{L}}(y)$,
\begin{eqnarray}
C(x-y) \propto \langle \bar \chi \chi(x) \rho_{\mathrm{L}}(y) \rangle_u
-\langle \bar \chi \chi\rangle_u
\langle \rho_{\mathrm{L}} \rangle_u,
\end{eqnarray}
as the function of $|x-y|$,
for $24^3 \times 24$ at $\beta$=6.0 
and $32^3 \times 32$ at $\beta$=6.2.
In both lattices, 
the bare quark masses of $m = 0.02, 0.015$, and $0.01$ 
are used, and the correlation function $C(x-y)$ 
is normalized to be unity at $|x-y|=0$ at each $m$. 
Within the error bar, the correlation function $C(x-y)$ seems to be a single-valued function of $|x-y|$, and  
no significant $m$-dependence of the correlation function is found in this bare quark-mass region. 

It is likely that 
the correlation function $C(x-y)$ 
monotonically decreases with the distance $r\equiv |x-y|$ 
and almost vanishes for large 
$r$ such as $r \gtrsim 0.5\, {\rm fm}$, 
and thus 
a strong correlation between 
the local chiral condensate and the monopole density is quantitatively clarified.

From these lattice QCD results, 
we conclude that there exists a direct clear local correlation between monopoles and the chiral condensate.

\section{Summary and conclusion}
We have studied spontaneous chiral-symmetry breaking in SU(3) QCD
in terms of the dual superconductor picture for quark confinement in the MA gauge, 
using lattice QCD Monte Carlo simulations 
with four types of different lattices. 
In the MA gauge, there topologically appear color-magnetic monopoles, 
which would be responsible to quark confinement. 

First, in the confinement phase, 
we have found Abelian dominance and monopole dominance 
for the chiral condensate in the chiral limit, 
using the two different methods of 
i) the Banks-Casher relation with the Dirac eigenvalue spectral density and 
ii) finite quark-mass calculations with the quark propagator and its chiral extrapolation. 
We have also found that bare-quark mass dependence of the chiral condensate 
is fairly reduced in Abelian-projected QCD and the monopole part.
In the high-temperature doconfined phase, 
the chiral restoration is observed
also for the Abelian and the monopole sectors. 

Second, we have investigated local correlation between the chiral condensate and color-magnetic monopoles, and have found that the chiral condensate takes a quite large value near the monopoles in Abelian-projected QCD.

Here, let us consider the physical origin of the correlation 
between chiral symmetry breaking 
and monopoles in terms of the magnetic catalysis. 
In Abelian gauge theories, chiral symmetry breaking is generally enhanced 
in the presence of a strong magnetic field, 
which is called the magnetic catalysis~\cite{Suganuma:1990nn,Klevansky:1992qe,Gusynin:1995nb}. 
In the MA gauge, infrared QCD resembles an Abelian gauge theory with monopoles,  
which accompany a strong color-magnetic field around them.  
Therefore, as an interesting possibility, 
the strong magnetic field around the monopoles enhances chiral symmetry breaking 
also in this Abelian gauge theory. 


As a future study, more detailed analysis on the local correlation would be desired
to determine what is the direct trigger of the enhancement of the local chiral condensate,
that is, the magnetic fields around monopoles, the presence of monopoles itself, or something else.
It is also meaningful and important 
to investigate the effect of dynamical quarks using full QCD.

\begin{acknowledgements}
H.S. is supported in part by the Grants-in-Aid for
Scientific Research [19K03869] from Japan Society for the Promotion of Science.
Most of numerical calculations have been performed on NEC SX-ACE and OCTOPUS at Osaka University, and Yukawa-21 at YITP, Kyoto University.
We have used PETSc and SLEPc to solve linear equations and eigenvalue problems for the Dirac operator, respectively 
~\cite{petsc-web-page,petsc-user-ref,petsc-efficient,Hernandez:2005:SSF}.
\end{acknowledgements}
\bibliography{correlation.bib}
\end{document}